\documentclass[
  twoside,
  twocolumn,
  aps,
  10pt,
  prb,
  preprintnumbers,
  floatfix,
  showpacs,
  citeautoscript,
  superscriptaddress,
]{revtex4-1}

\usepackage{amsmath}
\usepackage{amssymb}
\usepackage{graphicx}
\usepackage{tabularx}
\usepackage{dcolumn}
\usepackage{units}
\usepackage{hyperref}
\usepackage{multirow}
\usepackage{threeparttable}
\usepackage[export]{adjustbox}

\newcommand{\sect}[1]{Sect.~\ref{#1}}
\newcommand{\fig}[1]{Fig.~\ref{#1}}

\newcommand{\eq}[1]{Eq.~\eqref{#1}}

\newcommand{\tab}[1]{Table~\ref{#1}}

\renewcommand{\tensor}[1]{\ensuremath\mathbf{#1}}

\renewcommand{\vec}[1]{\ensuremath\boldsymbol{#1}}
\renewcommand{\epsilon}[0]{\varepsilon}

\newcommand{\BGG}{Ba$_8$Ga$_{16}$Ge$_{30}$}
\newcommand{\BGS}{Ba$_8$Ga$_{16}$Si$_{30}$}
\newcommand{\BAG}{Ba$_8$Al$_{16}$Ge$_{30}$}
\newcommand{\BAS}{Ba$_8$Al$_{16}$Si$_{30}$}
\newcommand{\BGGx}{Ba$_8$Ga$_{x}$Ge$_{46-x}$}
\newcommand{\BGSx}{Ba$_8$Ga$_{x}$Si$_{46-x}$}

\newcommand{\BASx}{Ba$_8$Al$_{x}$Si$_{46-x}$}

\newcommand{\myscale}{0.65}

\newcommand{\phys}{
  Chalmers University of Technology,
  Department of Physics,
  Gothenburg, Sweden
}
\newcommand{\chem}{
  Chalmers University of Technology,
  Department of Chemistry and Chemical Engineering,
  Gothenburg, Sweden
}

\begin{document}

\title{
Thermal conductivity in intermetallic clathrates:
A first principles perspective
}

\author{Daniel O. Lindroth}\affiliation{\phys}
\author{Joakim Brorsson}\affiliation{\chem}
\author{Erik Fransson}\affiliation{\phys}
\author{Fredrik Eriksson}\affiliation{\phys}
\author{Anders Palmqvist}\affiliation{\chem}
\author{Paul Erhart}\email{erhart@chalmers.se}\affiliation{\phys}

\begin{abstract}
Inorganic clathrates such as \BGGx{} and \BGSx{} commonly exhibit very low thermal conductivities.
A quantitative computational description of this important property has proven difficult, in part due to the large unit cell, the role of disorder, and the fact that both electronic carriers and phonons contribute to transport.
Here, we conduct a systematic analysis of the temperature and composition dependence of low-frequency modes associated with guest species in \BGGx{} and \BASx{} (``rattler modes''), as well as of thermal transport in stoichiometric \BGG{}.
To this end, we account for phonon-phonon interactions by means of temperature dependent effective interatomic force constants (TDIFCs), which we find to be crucial in order to achieve an accurate description of the lattice part of the thermal conductivity.
While the analysis of the thermal conductivity is often largely focused on the rattler modes, here, it is shown that at room temperatures modes with $\hbar\omega\gtrsim\unit[10]{meV}$ account for 50\%\ of lattice heat transport.
Finally, the electronic contribution to the thermal conductivity is computed, which shows the Wiedemann-Franz law to be only approximately fulfilled.
As a result, it is crucial to employ the correct prefactor when separating electronic and lattice contributions for experimental data.
\end{abstract}

\maketitle

\section{Introduction}

Thermoelectric materials enable the extraction of electrical power from a thermal gradient, as well as the reverse process, cooling through electrical power \cite{Row05, SnyTob08}.
As a result these materials are interesting for applications such as power generation in remote locations, waste heat recuperation, and active cooling.
Specifically, in the high-temperature region, which is of interest for example with regard to waste heat recuperation from combustion processes, inorganic clathrates are among the most efficient thermoelectric materials \cite{CohNolFes99, IvePalCox00} with studies reporting figure-of-merit ($zT$) values above one \cite{SarSvePal06, TobChrIve08}.

\begin{figure}[b]
  \centering
  \includegraphics[width=0.95\columnwidth, left]{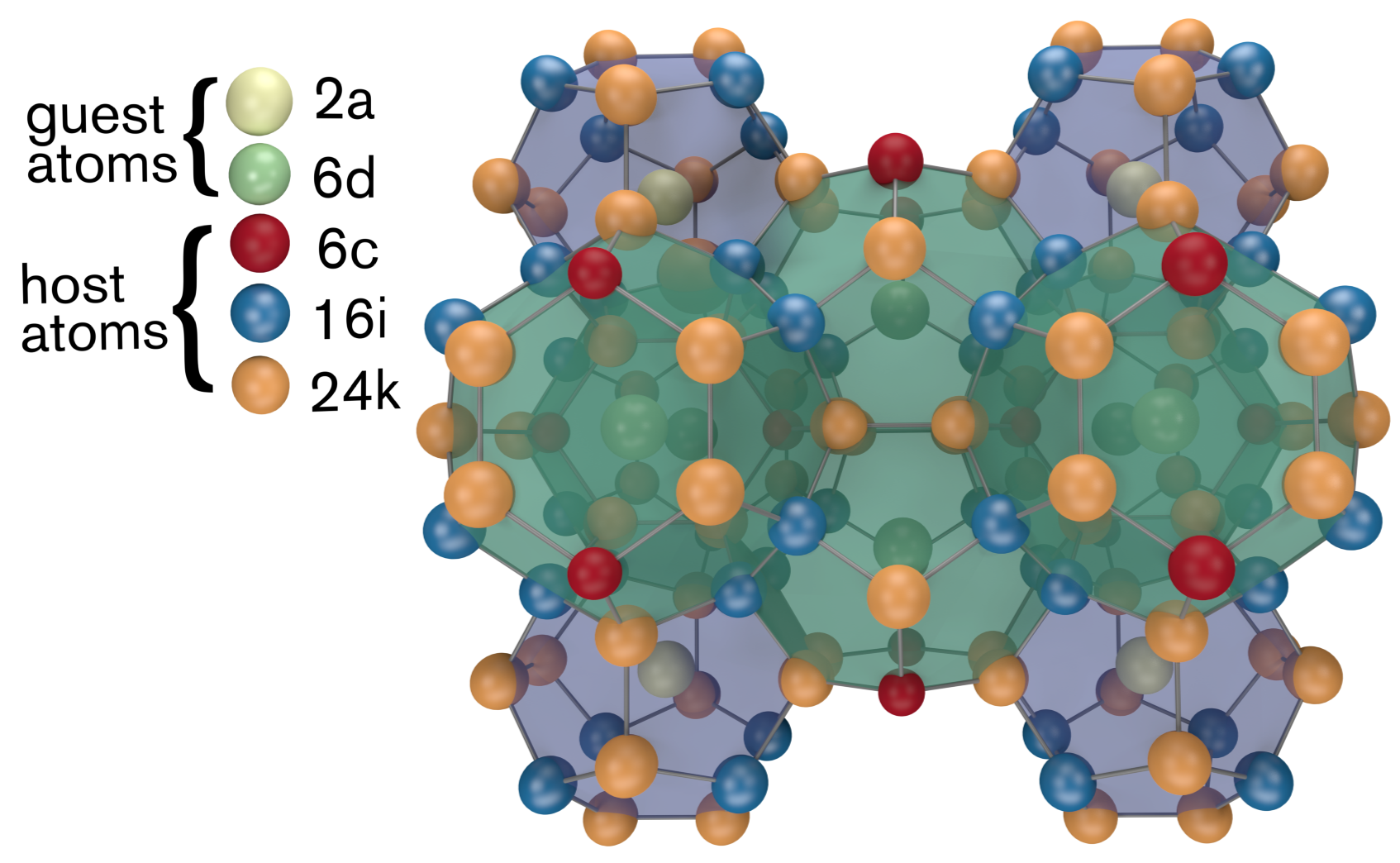}
  \caption{
    Crystal structure of type I clathrates.
    The guest species (Ba) occupies Wyckoff sites of type $2a$ and $6d$, while the host species (Ga, Ge) occupy Wyckoff sites of type $6c$, $16i$, and $24k$.
  }\label{fig:structure}
\end{figure}

Clathrates are chemical substances with a defined lattice structure that can trap atomic or molecular species \cite{IUPAC, MosSmiTav09}.
For thermoelectric applications one usually considers inorganic clathrates, examples of which include compounds such as \BGG{} or $\text{Sr}_8\text{Ga}_{16}\text{Sn}_{30}$ \cite{Rog05, SheKov11}.
Here, the earth alkaline atoms act as guest species that occupy the cages provided by the host structure, where the latter is most commonly composed of elements from groups 13 and 14.
In the present paper, we focus on \BGG, which belongs to space group P$m\bar{3}n$ (international tables of crystallography number 223) and features two smaller and six larger cages per unit cell (\fig{fig:structure}) \cite{Rog05, ChrJohIve10, SheKov11}.
\BGG{} has been investigated extensively both experimentally \cite{SalChaJin01, BryBlaMet02, SarSvePal06, ChrLocOve06, TobChrIve08, CedSarSny09, TakSueNak14} and theoretically \cite{BlaLatBry01, BlaBryLat01, BryBlaMet02, MadSchBla03, MadSan05, AngLinErh16, AngErh17, TadGohTsu15, TadTsu18}, especially because of its promising thermoelectric properties.

Generally these materials exhibit a very low thermal conductivity comparable to that of glasses \cite{CohNolFes99, DonSanMyl01, TakSueNak14}, which is crucial for their good thermoelectric performance, and accordingly clathrates are commonly regarded as realizations of the ``phonon glass-electron crystal'' concept \cite{Row95}.
This behavior can be attributed to the ``rattler''-like atomic motion of the guest species, which results from their relatively small size compared to the host cage \cite{MadSan05, ChrAbrChr08}.

While experimental measurements agree with respect to the general magnitude of the thermal conductivity in clathrates they exhibit some noticeable differences (illustrated for the case of \BGG{} in \fig{fig:lattice-thermal-conductivity} below).
These variations can originate from several factors including for example sample preparation, thermal emission, and the general difficulties associated with measuring small thermal conductivities.
Furthermore, as one is often interested in separating out the contribution to the thermal conductivity from lattice vibrations, one must make assumptions with respect to the electronic contribution, which adds another source of uncertainty.
In this situation, predictive modeling on the basis of first-principles calculations cannot only provide very valuable insight but also guide the development of new materials.
Compared to previous studies on \BGG{} \cite{TadGohTsu15, TadTsu18}, we provide a comprehensive comparison with experimental data for both vibrational spectrum and thermal conductivity, as well as an analysis of the composition dependence of the rattler modes and the electronic contribution to the thermal conductivity.
We furthermore emphasize the importance of accounting for the temperature dependence of phonon frequencies in order to obtain a sensible description of the thermal conductivity in this type of materials.

Here, we present a comprehensive study of the thermal conductivity in \BGG{} as a prototypical clathrate, which combines Boltzmann transport theory with first-principles calculations within the framework of density functional theory.
We address separately the electronic and vibrational contributions, account for finite-temperature effects on vibrational frequencies and lifetimes, consider the impact of the exchange-correlation functional, and conduct a careful comparison with experimental data.
It is demonstrated that while the Wiedemann-Franz law provides a reasonable approximation to the electronic thermal conductivity, it is crucial to use the appropriate pre-factor, an aspect that is often overlooked in the analysis of experimental data.

Furthermore it is shown that the rattler-mode frequencies calculated in the static (zero temperature) limit systematically underestimate the experimental data, which leads to a striking underestimation of the lattice thermal conductivity.
To overcome this limitation one must take into account phonon renormalization, which is accomplished effectively by using temperature dependent force constants.
The resulting model reproduces the experimentally measured temperature dependence of the rattler-mode frequencies and leads to an increase of the thermal conductivity by more than a factor of two, which brings the predicted data in good agreement with experimental data.
Finally, we provide a careful examination of the contributions to the thermal conductivity, which shines light on the ``phonon glass'' picture and reveals that more modes contribute to the thermal conductivity than previously assumed.

The remainder of this paper is organized as follows:
The next section provides an overview of the methodological aspects of this work including computational details as well as a review of the most important relations from Boltzmann transport theory.
The first part of the Results and Discussion section then focuses on the suitability of different exchange-correlation functionals to describe the clathrate structures.
This is followed by an examination of the phonon dispersion, especially the rattler modes, emphasizing their temperature dependence as well as the effect of order and composition.
This sets up a comprehensive analysis of the vibrational and electronic contributions to the thermal conductivity.
Finally, we summarize the key results and conclusions in \sect{sect:conclusions}.

\section{Methodology}
\label{sect:methodology}

\subsection{Thermal conductivity}

The thermal conductivity $\kappa$ in a solid chiefly comprises contributions from electronic carriers $\kappa_e$ and lattice vibrations (phonons) $\kappa_l$,
\begin{align}
    \kappa = \kappa_e + \kappa_l.
\end{align}
In the following we outline the approach taken to compute the two terms in the above equation.

\subsubsection{Lattice thermal conductivity}

The lattice (phononic) contribution $\vec{\kappa}_l$ to the thermal conductivity can be computed by solving the phonon Boltzmann transport equation (BTE) \cite{Zim60}.
In the present treatment, we limit our analysis to the framework of the relaxation time approximation (RTA) of the phonon BTE, in which the lattice thermal conductivity is
\begin{align}
  \vec{\kappa}_l &= \frac{1}{\Omega}
  \sum_{i\vec{q}} g_{\vec{q}}  \vec{v}_{i\vec{q}} \otimes \vec{v}_{i\vec{q}} \tau_{i\vec{q}} c_{i\vec{q}}.
  \label{eq:thermal-conductivity}
\end{align}
Here, $\Omega$ is the unit cell volume, $g_{\vec{q}}$ is the $\vec{q}$-point weight, and $\vec{v}_{i\vec{q}}= \nabla_{\vec{q}} \omega_{i\vec{q}}$ is the group velocity of mode $i$ at point $\vec{q}$ of the Brillouin zone with $\omega_{i\vec{q}}$ being the mode frequency.
Both phonon frequencies and group velocities can be derived from the second-order force constant matrix, which is given by the second derivative of the energy $E$ with respect to the atomic displacements $u_{\alpha}(il)$ \cite{Zim60}
\begin{align}
    \Phi_{\alpha\beta}(il,i'l') = \frac{\partial^2 E}{\partial u_{\alpha}(il)\partial u_{\beta}(i'l')},
\end{align}
where $\alpha$ and $\beta$ are Cartesian directions, $i$ is the site index relative to the unit cell basis, and $l$ an index enumerating the unit cells.
From the force constant matrix one can readily compute the dynamical matrix at any momentum vector $\vec{q}$,
\begin{align}
    D_{\alpha \beta}(jj',\vec{q}) = \frac{1}{\sqrt{m_jm_{j'}}}\sum_{l'}\Phi_{\alpha\beta}(j0,j'l')e^{i\vec{q}\cdot (r_{j'l'}-r_{j0})},
    \label{eq:dynamical-matrix}
\end{align}
where $m_j$ is the atomic mass of the species occupying site $j$.
Diagonalization of $D_{\alpha \beta}(jj',\vec{q})$ then yields normal modes and phonon frequencies $\omega_{i\vec{q}}$, from which the mode-specific heat capacity $c_{i\vec{q}}$ at temperature $T$ can be obtained via
\begin{align}
  c_{i\vec{q}} &= k_B \frac{x^2 \exp x}{{\left(1-\exp x\right)}^2}
  \quad\text{with}\quad
  x = \frac{\hbar\omega_{i\vec{q}}}{k_B T}.
  \label{eq:heat-capacity}
\end{align}

The relaxation time $\tau_{i\vec{q}}$, or phonon lifetime, which appears in \eq{eq:thermal-conductivity}, comprises contributions from different scattering processes including, e.g., phonon-phonon interaction, isotope mass variation, boundary scattering, alloying, and disorder.
Here, we consider phonon-phonon interaction and isotope mass variation.
According to the most simple approximation, known as Matthiessen's rule, the different scattering processes are assumed to be independent, i.e. their scattering rates (or inverse lifetimes) are additive
\begin{align}
    \tau^{-1}_{i\vec{q}} = \tau^{-1}_{ph-ph,i\vec{q}} + \tau^{-1}_{iso,i\vec{q}}.
    \label{eq:matthiesen}
\end{align}
In the present work scattering due to isotope mass variation ($\tau_{iso,i\vec{q}}$) has been treated according to second order perturbation theory \cite{Tam83} whereas the contribution due to phonon-phonon scattering ($\tau^{-1}_{ph-ph,i\vec{q}}$) was treated at the level of first-order perturbation theory \cite{Zim60}, which requires knowledge of not only the second but also third-order interatomic force constants (IFCs) \cite{LiCarKat14, LiCarMin13}.
As detailed in \sect{sect:computational-details}, we computed IFCs both in the static (0\,K) limit using the finite displacement method and from molecular dynamics (MD) simulations.

\subsubsection{Electronic thermal conductivity}

In the relaxation time approximation (RTA) to the linearized Boltzmann transport equation (BTE) the electronic contribution to the thermal conductivity $\vec{\kappa}_e$ is given by
\begin{align}
    \vec{\kappa}_e = \vec{\kappa}^0 - \vec{S}^2\vec{\sigma} T
    \label{eq:electronic_contribution_thermal_conductivity}
\end{align}
with \cite{Zim60, BlaMolKre99, MadSin06}
\begin{align}
  \vec{\sigma} &= \frac{2 e^2}{\Omega}
   \sum_{i\vec{k}} g_{\vec{k}} \vec{v}_{i\vec{k}} \otimes \vec{v}_{i\vec{k}} \tau_{i\vec{k}}
  {\left( \frac{\partial f}{\partial \epsilon} \right)}_{\epsilon=\epsilon_{i\vec{k}}}\label{eq:electrical-conductivity}
  \\
  \tensor{S} &= \frac{\vec{\sigma}^{-1}}{e T}
   \sum_{i\vec{k}} g_{\vec{k}} \vec{v}_{i\vec{k}} \otimes \vec{v}_{i\vec{k}} \tau_{i\vec{k}}
  \left[ \epsilon_{i\vec{k}} - \mu_e \right]
  {\left(\frac{\partial f}{\partial \epsilon}\right)}_{\epsilon=\epsilon_{i\vec{k}}}
  \label{eq:seebeck-coefficient}
  \\
  \vec{\kappa}^0 &= \frac{e}{T\Omega}\sum_{i\vec{k}}g_{\vec{k}}
  \vec{v}_{i\vec{k}} \otimes \vec{v}_{i\vec{k}}\tau_{i\vec{k}}
  {\left[ \epsilon_{i\vec{k}} - \mu_e \right]}^2
  {\left(\frac{\partial f}{\partial \epsilon}\right)}_{\epsilon=\epsilon_{i\vec{k}}}
  \label{eq:electronic-thermal-conductivity}.
\end{align}
Here, $\Omega$ is the unit cell volume, $g_{\vec{k}}$ is the $ \vec{k}$-point weight, $i$ refers to the band index, $\tau_{i\vec{k}}$ is the mode and momentum dependent lifetime, $\vec{v}_{i\vec{k}}=\hbar^{-1} \partial \epsilon_{i\vec{k}}/\partial \vec{k}$ is the group velocity, $f$ is the occupation function, and $\mu_e$ is the electron chemical potential.

We have previously studied the electronic conductivity $\vec{\sigma}$ and the Seebeck coefficient $\tensor{S}$ for \BGG{} and conducted a systematic comparison with experiment \cite{AngLinErh16}.
Using a charge carrier concentration of $n_e = 3 \times 10^{20}\,\text{cm}^{−3}$ and a mode and momentum-independent effective lifetime model with $\tau_\text{eff}=\tau_{300}(300\,\text{K}/T)^{1/2}$ we were able to achieve very good agreement with experimental data, and accordingly this approach is also adopted in the present work.

\subsection{Computational details}
\label{sect:computational-details}

\subsubsection{General}

DFT calculations were performed using the projector augmented wave method \cite{Blo94, *KreJou99} as implemented in the Vienna \emph{ab initio} simulation package (\textsc{vasp}) \cite{KreFur96b}.
To assess the sensitivity of our results to the treatment of exchange-correlation effects, we used both the PBE functional \cite{PerBurErn96} and the van der Waals density functional method \cite{BerCooLee15} with consistent exchange (vdW-DF-cx) \cite{BerHyl14b} as implemented in \textsc{vasp} \cite{KliBowMic11, Bjo14}.

The plane-wave energy cutoff energy was set to 243\,eV (\BGG), 312\,eV (\BAG), and 319\,eV (\BAS, \BGS), respectively, in calculations at fixed volume and cell shape.
For cell shape relaxations the plane-wave energy cutoff was increased by 30\%.
A Gaussian smearing with a width of 0.1\,eV was used throughout.
Structural relaxations were performed using a $\Gamma$-centered $3\times3\times3$ $\vec{k}$-point mesh until the residual forces were below 10\,meV/\AA\ and absolute stresses were below 0.1\,kbar.

\subsubsection{Vibrational spectra and lattice thermal conductivity}

The static second and third-order IFCs as well as the thermal conductivity were computed using the \textsc{shengBTE} \cite{LiMinLin12, LiCarKat14, LiLinBro12} and \textsc{phonopy} \cite{TogObaTan08} codes.
Calculations were carried out using $2\times2\times2$ supercells (432 atoms) and in the case of the third-order IFCs included displacements up to the fifth neighbor shell.
The Brillouin zone was sampled using a $9\times9\times9$ $\vec{q}$-point mesh and a smearing parameter of $\sigma=0.01$.
There was no indication of any significant difference between the results within the framework of RTA-BTE, and the fully converged solution to the BTE, hence the full set of computations was limited to the RTA.
The second-order IFCs obtained in this process were also used to model the thermal expansion within the quasi-harmonic approximation (QHA).

As will be shown below, using the IFCs obtained in the static (0\,K) limit to predict the thermal conductivity leads to a substantial underestimation.
We therefore also determined effective temperature dependent IFCs using an approach similar to the one described in Ref.~\onlinecite{HelAbr13}.
To this end, we carried out first-principles molecular dynamics (MD) simulations in the canonical ensemble at temperatures of 100, 200, 300, and 600\,K.
We employed primitive 54-atom cells, which were sampled using a $\Gamma$-centered $3\times3\times3$ $\vec{k}$-point mesh.
The equations of motion were integrated for a total of about 5,500 time step using a time step of 5\,fs.
After discarding the first 1,000 steps for equilibration, about 180 snap shots at a spacing of 25 MD steps were used for training, by least-squares fitting temperature dependent interatomic force constants (TDIFCs) using our in-house \textsc{hiphive} code \cite{EriFraErh19}.
Finally, \textsc{shengBTE} was used to calculate the thermal conductivity from the resulting IFCs.

\subsubsection{Vibrational spectra of non-stoichiometric compounds}
\label{sect:method-frequency-composition}

Additional calculations of the vibrational spectra were carried out for \BGGx{} and \BASx{} for $14\leq x\leq 18$.
To this end, we employed 54-atom cells and structures obtained previously by Monte Carlo simulations that are representative of the actual chemical order in the material.
In total data was obtained for 20 structures per composition, equivalent to 200 configurations in total.

\subsubsection{Electronic contribution to the thermal conductivity}

To calculate the electronic contribution to the thermal conductivity we considered both the chemically ordered ground state structure and the chemically disordered structures obtained by Monte Carlo simulations \cite{AngLinErh16}.
The latter configurations are representative of the actual chemical (dis)order in the material at 600, 900, and 1200\,K as described in detail in Ref.~\onlinecite{AngLinErh16}.
Results were averaged over five structures per temperature.
The wave function of the fully relaxed structures were converged using a $\Gamma$-centered $4\times4\times4$ $\vec{k}$-point mesh, which was followed by a non-self-consistent computation of the eigenenergy spectra on a $\Gamma$-centered $20\times20\times20$ mesh.
The terms in \eq{eq:electronic_contribution_thermal_conductivity} were subsequently computed using the \textsc{BoltzTrap} code \cite{MadSin06}.

\section{Results and discussion}

\subsection{Structure and thermal expansion}
\label{sect:clathrate-structure}

\begin{figure}[b]
  \centering
  \includegraphics[scale=\myscale, left]{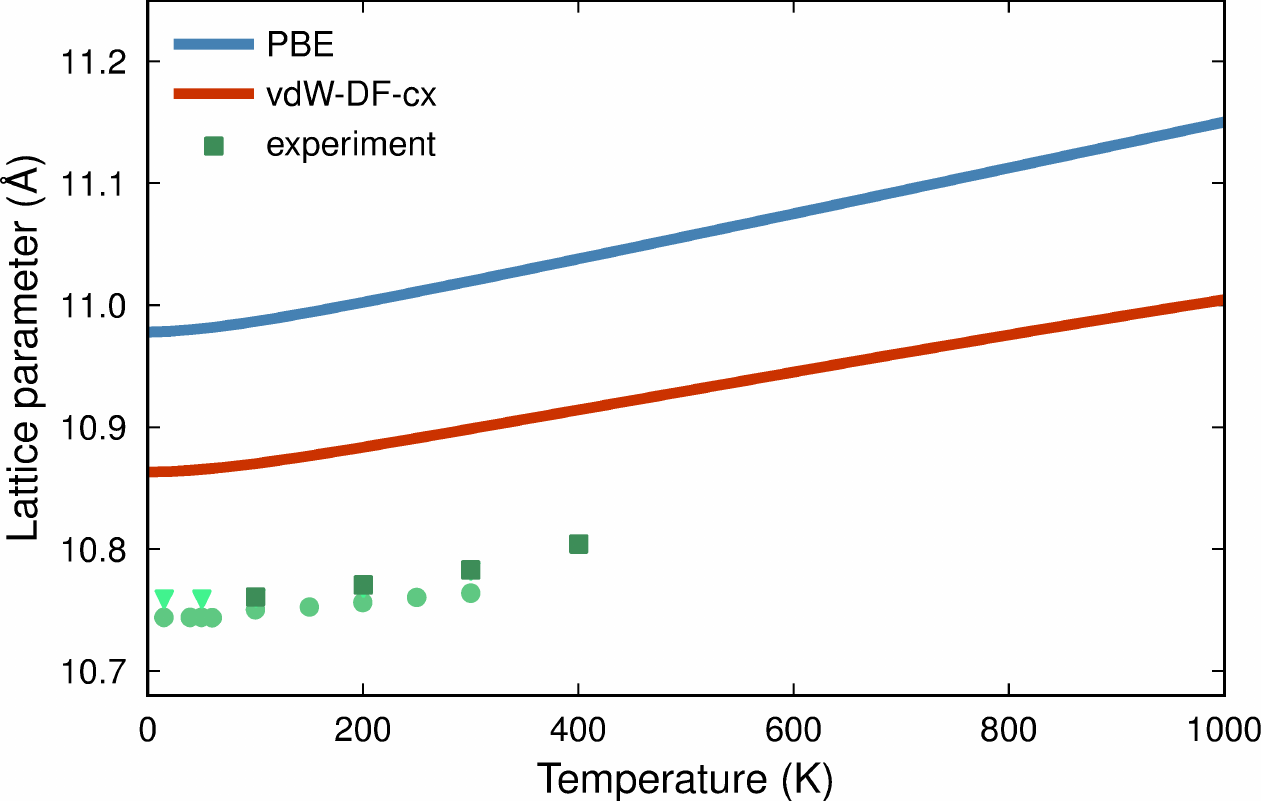}
  \caption{
    Temperature dependence of the lattice parameter for \BGG{} obtained within the quasi-harmonic approximation.
    Experimental data are from Ref.~\onlinecite{ChrJohIve10}.
  }
  \label{fig:lattice-constant-BGG}
\end{figure}

An accurate description of the vibrational properties is important for modeling the thermal conductivity.
While the PBE functional has been used extensively in the past in electronic structure calculations of inorganic clathrates (see e.g., Refs.~\onlinecite{MadSan05, TadGohTsu15}), it is known to underestimate the bond stiffness (see e.g., Ref.~\onlinecite{GhaErhHyl17}).
In the present work, we therefore also considered the vdW-DF-cx method (see \sect{sect:computational-details}), which has been found to yield an excellent description of finite temperature properties for other materials \cite{GhaErhHyl17, LinErh16}.

As a first step in assessing the performance of these functionals we determined the temperature dependence of the lattice structure of the four stoichiometric compounds \BGG, \BGS, \BAG, and \BAS\ based on the ordered ground state (0\,K) structure determined in Ref.~\onlinecite{AngErh17}.
In this context it must be noted that it is not always possible to find experimental data for the lattice parameters for the stoichiometric compounds.
This applies in particular to the Si-based clathrates \cite{ChrJohIve10, AnnYamNak12, TsuRouZev11, NagMugNak12}, presumably because the solubility limit is about $x=15$ \cite{AnnYamNak12, TsuRouZev11} for both Ga in \BGSx\ and Al in \BASx.
The lattice parameters that correspond to $x=16$ have been estimated by performing a linear least-squares fit for each data set.
It shall also be noted that the experimentally determined lattice parameters and compositions can vary markedly depending on the measurement technique \cite{AnnYamNak12, ChrIve07, ChrJohIve10}.

\begin{table}[b]
  \centering
  \begin{threeparttable}[t]
  \caption{
      Finite temperature lattice parameters from calculation and experiment.
      The values in brackets denote the lattice parameters obtained without taking into account zero-point vibrations.
      In the calculations the temperature dependence was described at the level of the quasi-harmonic approximation.
  }
  \label{tab:structural-parameters}
  \begin{tabularx}{\columnwidth}{lX*{5}d*{2}d}
    \toprule
    \multicolumn{2}{l}{Material}
    & \multicolumn{5}{c}{Calculation}
    & \multicolumn{2}{c}{Experiment}
    \\

    \colrule{}
    \\[-9pt]

    &
    & \multicolumn{2}{c}{0\,K}
    && \multicolumn{1}{c}{300\,K}
    && \multicolumn{2}{c}{300\,K}
    \\

    \cline{3-4}
    \cline{6-6}
    \cline{8-9}
    \\[-6pt]

    \multicolumn{7}{l}{\BGG} & \multicolumn{1}{c}{\text{Ref. \onlinecite{OkaKisTan06}}} & \multicolumn{1}{c}{\text{Ref. \onlinecite{ChrLocOve06}}} \\
    PBE && 10.98  & (10.96)  &&  11.02 && 10.76\tnote{a} & 10.80\tnote{a} \\
    vdW-DF-cx && 10.86  & (10.85)  &&  10.90 && & \\
    \multicolumn{7}{l}{\BGS} & \multicolumn{1}{c}{\text{Ref. \onlinecite{ChrJohIve10}}} & \multicolumn{1}{c}{\text{Ref. \onlinecite{AnnYamNak12}}} \\
    PBE && 10.68  & (10.66)  &&  10.71 && 10.54\tnote{b} & 10.57\tnote{b} \\
    vdW-DF-cx && 10.60  & (10.58)  &&  10.62 && & \\
    \multicolumn{7}{l}{\BAS} & \multicolumn{1}{c}{\text{Ref. \onlinecite{NagMugNak12}}} & \multicolumn{1}{c}{\text{Ref. \onlinecite{TsuRouZev11}}} \\
    PBE && 10.74  & (10.72)  &&  10.76 && 10.64\tnote{b} & 10.65\tnote{b} \\
    vdW-DF-cx && 10.67  & (10.65)  &&  10.69 && & \\
    \multicolumn{7}{l}{\BAG} & \multicolumn{1}{c}{\text{Ref. \onlinecite{ChrIve07}}} & \multicolumn{1}{c}{\text{Ref. \onlinecite{RodSarRos10}}} \\
    PBE && 11.01  & (10.99)  &&  11.04 && 10.85\tnote{a} & 10.88\tnote{b} \\
    vdW-DF-cx && 10.91  & (10.90)  &&  10.94 && & \\
    \botrule
  \end{tabularx}
  \begin{tablenotes}
  \item[a] Interpolated
  \item[b] Extrapolated
  \end{tablenotes}
  \end{threeparttable}
\end{table}

Both functionals overestimate the lattice parameter compared to experiment, with PBE always giving the higher estimate (\tab{tab:structural-parameters} and \fig{fig:lattice-constant-BGG}).
Overall the agreement achieved by the vdW-DF-cx calculations is very good with an average deviation of 0.6\%\ (1.5\%\ for PBE).

It should be noted that chemical disordering, which is generally present in these compounds \cite{ChrJohIve10}, has an effect on the lattice parameter.
Based on our earlier analysis \cite{AngErh17}, one can assume that as the material is cooled down after synthesis the chemical order is frozen in at a temperature of about 600\,K.
At this temperature the lattice parameter in \BGG\ has been predicted to be increased by $0.028\,\text{\AA}$ relative to the ground state structure, decreased by about $0.028\,\text{\AA}$ in the case of \BGS\, and relatively unchanged in the case of \BAG\ and \BAS.
These contributions, however, barely affect the agreement with experiment and leave the average errors unchanged.

\subsection{Phonon dispersion in the static \texorpdfstring{(0\,K)} limit}

\begin{figure*}
    \includegraphics[scale=\myscale,center]{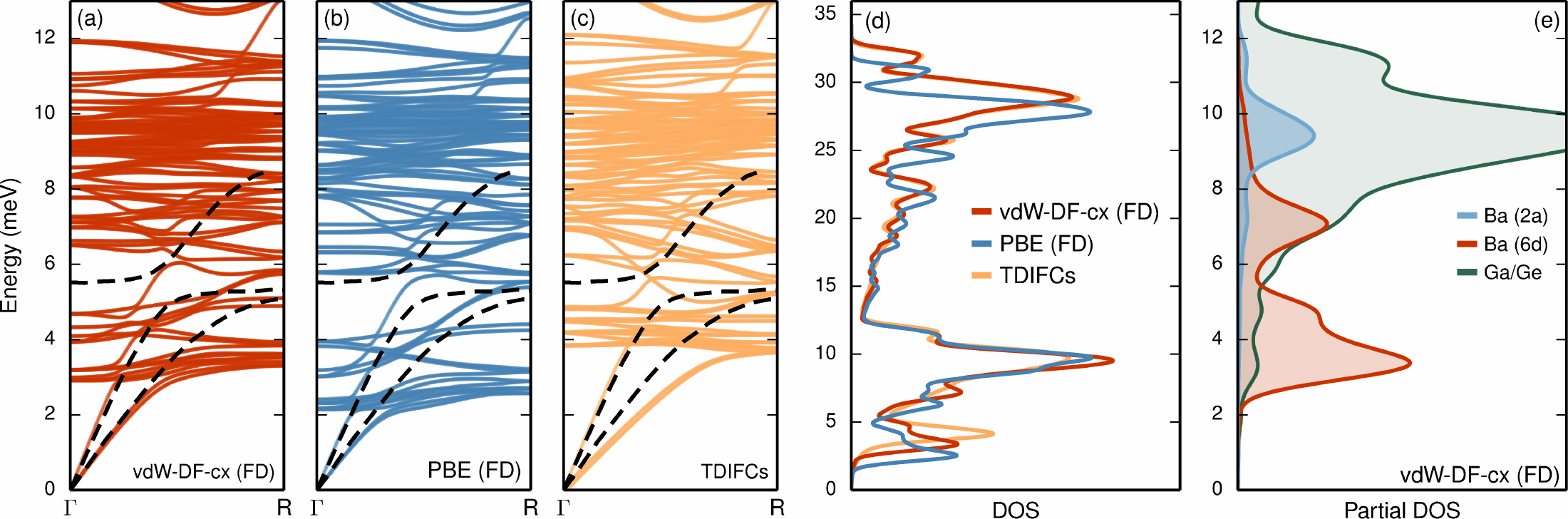}
    \caption{
        (a-c) Phonon dispersion of \BGG{} showing the low frequency region along the $\Gamma$-\textsf{R} direction derived (a,b) from IFCs obtained in the static (0\,K) limit by the finite-displacement (FD) method and (c) from temperature dependent force constants (TDIFCs) corresponding to a temperature of 300\,K.
        Black dashed lines indicate the result of a simple spring model fitted to experimental data \cite{ChrAbrChr08}.
        The model underestimates the splitting at the zone boundary close to 5\,meV, which is actually approximately 1\,meV.
        (d) Total phonon densities of states.
        (e) Partial densities of states showing the contributions from Ba on 2a Wyckoff sites (blue line), Ba on 6d sites (red line) and contributions from the Ga/Ge cage structure (green).
    }
    \label{fig:ph_dispersions_dos}
\end{figure*}

Due to the large mass of the Ba atoms as well as their weak coupling to the host structure, the associated rattler modes show up as low-frequency optical modes in the phonon spectrum (\fig{fig:ph_dispersions_dos}).
They appear at higher frequencies in the vdW-DF-cx calculations [\fig{fig:ph_dispersions_dos}(a)] than in the case of the PBE functional [\fig{fig:ph_dispersions_dos}(b)] as expected based on the known ``softness'' of the latter.
For both functionals one observes the phonon modes in the zero temperature limit to be lower than the experimentally measured frequencies [black dashed lines in \fig{fig:ph_dispersions_dos}(a-c)] \cite{ChrAbrChr08}.

\subsection{Phonon dispersion at finite temperatures}

\begin{figure*}
    \centering
    \includegraphics[scale=\myscale, left]{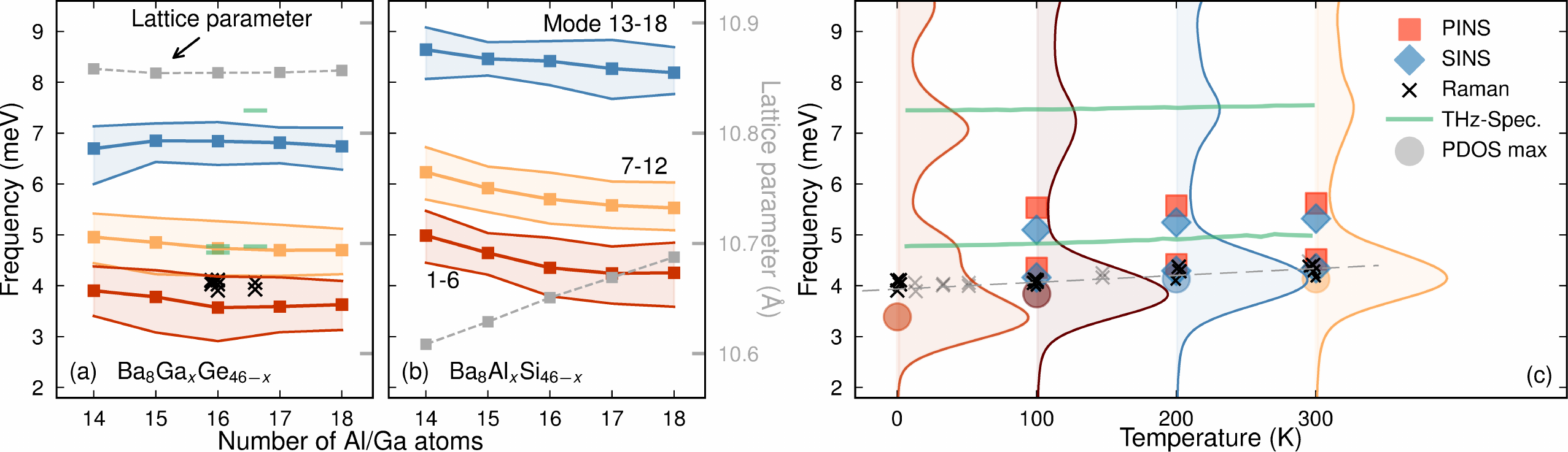}
    \caption{
        (a,b) Composition dependence of the lowest lying rattler modes in (a) \BGGx\ and (b) \BASx\ along with the variation of the lattice constant (in gray).
        Specifically, the red, yellow and blue lines correspond to the average frequencies for modes 1--6, 7--12 and 13--18, respectively.
        The standard deviations obtained by averaging over 20 representative configurations at each composition are indicated by shaded filled curves.
        (c) Temperature dependence of the partial density of states associated with Ba atoms on 6d Wyckoff sites (shaded filled curves, compare \fig{fig:ph_dispersions_dos}(e)) in comparison with experimental data from inelastic neutron scattering on powder (PINS) and single crystalline (SINS) samples \cite{ChrAbrChr08}, Raman Scattering \cite{ChrAbrChr08, TakHasOgi06, TakHasOgi08JPSJ, TakHasOgi08PRL, TakHasOgi10} and THz spectroscopy \cite{MorGosIwa09, IwaMorKus12, IwaKusHon13}.
    }
    \label{fig:raman_effective_pdos}
\end{figure*}

As will be shown below the thermal conductivity calculated on the basis of the static IFCs systematically and substantially underestimates experimental data (also see Ref.~\onlinecite{TadGohTsu15}).
As will become apparent in the analysis of the thermal conductivity (\sect{sect:lattice-thermal-conductivity}) lower frequencies of the rattler modes reduce the Brillouin zone volume corresponding to propagating modes, which translates into a lower thermal conductivity.
It is therefore a very relevant question to which extent phonon-phonon interactions affect the rattler mode frequencies.
In fact the low frequencies of the rattler modes imply that they are fully activated already at low temperatures and thus phonon-phonon interaction driven frequency shifts can already occur below room temperature.
This notion is supported by experimental data from both inelastic neutron and Raman scattering \cite{ChrAbrChr08} that reveals a notable temperature dependence of the rattler modes.

Using a series of temperature dependent interatomic force constants (TDIFCs, see \sect{sect:computational-details}), we therefore calculated the vibrational spectrum as a function of temperature.
The full phonon dispersion at 300\,K [\fig{fig:ph_dispersions_dos}(c)] does indeed reveal an upward shift of the lowermost optical branches.
A comprehensive comparison with experimental data [\fig{fig:raman_effective_pdos}(c)] demonstrates that the TDIFCs can also rather accurately reproduced both the absolute values and the temperature dependence of the rattler modes.
\footnote{
    We note that since only certain vibrational motions can be detected with Raman and THz spectroscopy techniques, it is possible to indirectly draw some conclusions regarding the symmetries of the modes.
    Specifically, those that are Raman active have either $T_{2g}$  or $E_{g}$ symmetries and represent vibrations parallel to the $[100]$  and  $[110]$ directions \cite{TakHasOgi10}.
    With THz spectroscopy, however, the only visible, guest atom modes are those that are infrared active, and have $T_{1u}$ symmetries \cite{MorGosIwa09}.
    Presumably, the lower and higher of these two modes correspond to motions perpendicular and parallel to the $[001]$ direction (the out-of-plane direction in \fig{fig:structure}), respectively.
}
As will be discussed below these effects are actually crucial for being able to predict correctly the thermal conductivity.

\subsection{Chemical composition and ordering}
\label{sect:results-frequencies-composition}

According to ex\-p\-eri\-men\-tal \cite{NatNag04, AviSueUme06a, AviSueUme06b, ChrIve07, MarWanNol08, MayTobSar09, TanKumJu09, TakHasOgi10, TsuRouZev11, NagMugNak12}, theo\-re\-ti\-cal \cite{AngErh17, TroRigDra17} as well as combined \cite{UemAkaKog08, RodSarRos10} studies, the structural and physical properties of, ternary, inorganic clathrates vary markedly with chemical composition.
Specifically, it has been shown that the displacement of the guest atom from the $6d$ site \cite{ChrIve07, AngErh17} and the associated vibrational frequencies\cite{NatNag04, TakHasOgi10} depend on the number of Al or Ga atoms per unit cell, in ternary compounds of the type Ba$_8\{$Al,Ga$\}_x\{$Si,Ge$\}_{46-x}$.
Moreover, experimental evidence suggests that the degree of off-centering, the frequencies of the lowest Raman active modes and the lattice thermal conductivity are correlated for compounds in the structurally similar quaternary system Sr$_8$Ga$_{16}$Si$_x$Ge$_{30-x}$ \cite{TakHasOgi08JPSJ, TakHasOgi08PRL}.

Given these results, it is reasonable to assume that the lattice thermal conductivity also varies to some degree with chemical composition.
We therefore computed the variation of the 18 lowest-frequency phonon modes, associated with the Ba atom at the $6d$ site \footnote{The ``center'' for each of the $3\times54=162$ phonon modes, in the form of the eigenvalues of the Hessian matrix, was taken as the atom that gave the largest contribution to the phonon density of states at that particular frequency.}, with the number of group-13 atoms not only in \BGGx{} but also \BASx{}, where the latter was included as it represents the limit of a host matrix made up of light elements.
Specifically, we extracted and averaged the $\Gamma$-point frequencies for 20 representative configurations (\sect{sect:method-frequency-composition}) for each composition in the range $14\leq x\leq 18$.
The modes naturally fall into three groups, with six modes in each [\fig{fig:raman_effective_pdos}(a,b)].
The splitting of the modes can be viewed as a consequence of the facts that (\emph{i}) the guest atom is not located at the immediate center of the cage, (\emph{ii}) the latter is shaped like a tetrakaidekahedron, and (\emph{iii}) the Al and Ga atoms are not necessarily symmetrically distributed between the framework sites \cite{AngErh17}.

For \BASx{} the phonon modes slightly soften with increasing $x$; a similar trend albeit even weaker can also be observed for the lower two groups in the case of \BGG.
This behavior correlates with the increase in the lattice constant, which is larger for \BASx{} than for \BGGx{} [gray lines in \fig{fig:raman_effective_pdos} (a,b)].
A larger lattice parameter implies that the size of the cages occupied by Ba atoms increases, which leads to weaker restoring forces and, hence, lower vibrational frequencies.
Overall one must conclude, however, that the phonon frequencies, associated with the vibrations of the Ba atom at the $6d$ Wyckoff, are relatively insensitive to the chemical composition since the difference between $x=14$ and $x=18$ is of the same magnitude as the spread of the frequencies.

\subsection{Thermal conductivity: lattice contribution}
\label{sect:lattice-thermal-conductivity}

\subsubsection{Comparison of scattering channels}

\begin{figure}
    \centering
    \includegraphics[scale=\myscale, left]{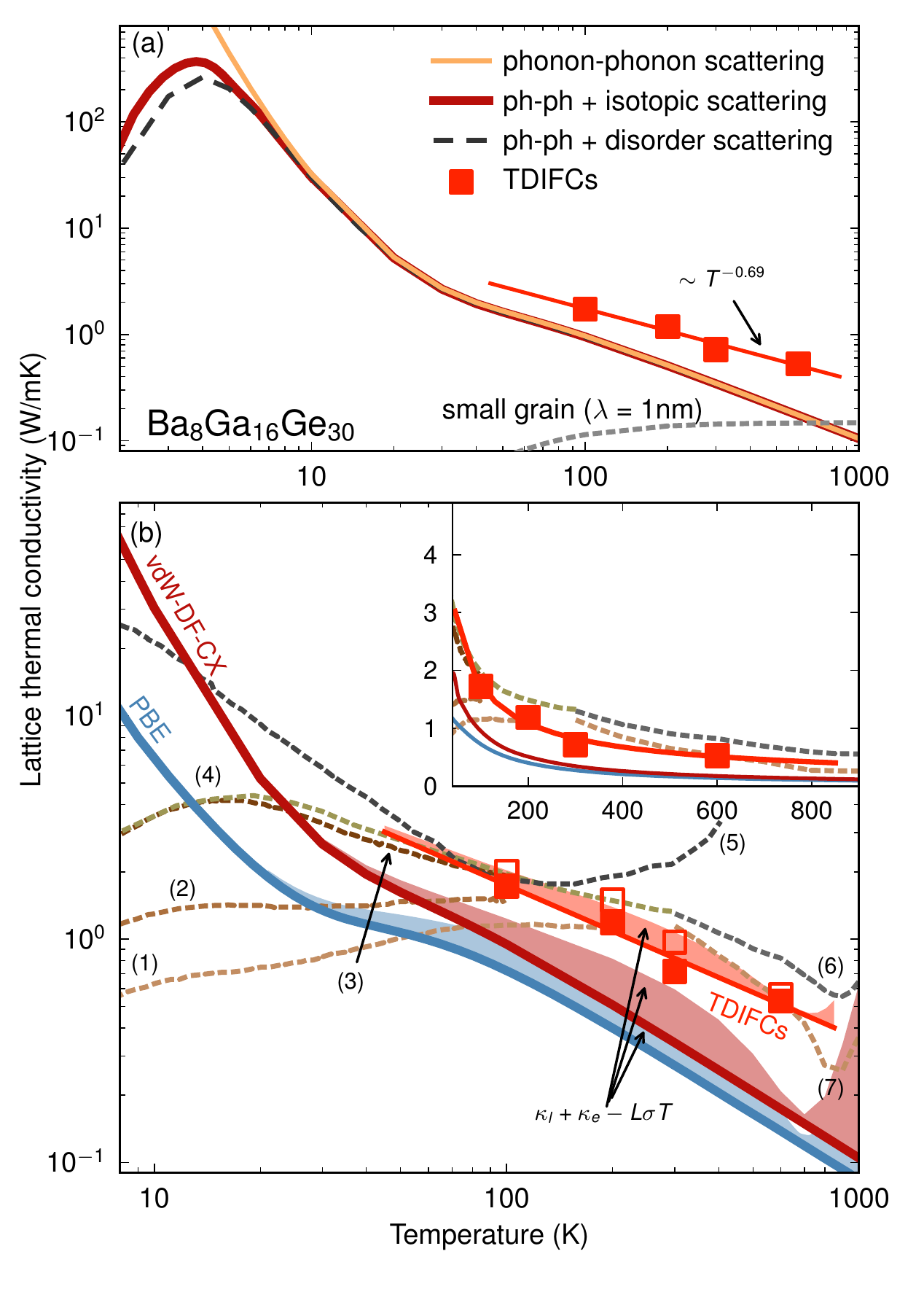}
    \caption{
        Lattice thermal conductivity $\kappa_l$ of \BGG{} as a function of temperature.
        (a) Comparison of $ \kappa_l$ due to different included scattering channels, calculated using IFCs achieved with the vdW-DF-cx functional, as well as a comparison to the computations using temperature dependent force constants (TDIFCs).
        (b) Comparison between computations, using IFCs from the vdW-DF-cx functional (red solid line), the PBE functional (solid blue line), temperature dependent force constants (squares) and experimental data sets (dashed lines marked by numbers).
        The inset shows the same data on a linear scale.
        The experimental data are from Refs.~\onlinecite{BenChrBry04} (1, p-type), \onlinecite{AviSueUme06b} (2, p-type; 3, n-type), \onlinecite{SalChaJin01} (4, n-type), \onlinecite{ChrJohSon09} (5, n-type), \onlinecite{HouZhoWan09} (6, 7).
        The authors of Ref.~\onlinecite{ChrJohSon09} (5) point out that due to the large surface to volume ratio, their measurements become unreliable above approximately 100\,K due to thermal emission.
        The shaded areas and open squares represent the difference between the electronic contribution to the thermal conductivity from BTT $\kappa_e$ and the Wiedemann-Franz law $L \sigma T$ with $L=2.0(k_B/e)^2$.
    }
    \label{fig:lattice-thermal-conductivity}
\end{figure}

Based on the analysis of the vibrational properties described in the previous sections we computed the lattice contribution to the thermal conductivity.
These calculations were carried out for the chemically ordered ground state structure of \BGG{} \cite{AngLinErh16, AngErh17} using both the PBE functional and the vdW-DF-cx method.

When limiting the analysis to phonon-phonon scattering, one obtains a strong variation with temperature that follows a $T^{-2}$ behavior at low temperatures and a $T^{-1}$ trend at temperatures $\gtrsim$ \unit[100]{K} [\fig{fig:lattice-thermal-conductivity}(a)].
Isotope scattering affects only the very low temperature region leading to a peak in the conductivity below 10\,K.

The chemical disordering at finite temperatures that was already alluded to above is inevitably associated with mass mixing, which affects both the frequencies via \eq{eq:dynamical-matrix} and the lifetimes in a way analogous to isotope scattering.
The effect on the frequencies is modest as evident from the analysis in \sect{sect:results-frequencies-composition}, that is to say the second-order force constants are hardly affected by chemical mixing on the host lattice, at least at a Ga:Ge ratio of 16:30.

By extension of \eq{eq:matthiesen}, mass disorder due to chemical mixing can, however, also affect the lifetimes.
Here, this effect was modeled analogous to isotope (mass) scattering using site occupancy factors obtained in previous simulations to compute the variance of the atomic masses \cite{AngErh17}.
The results show that the impact on the lifetimes is only discernible at very low temperatures [dashed black line in \fig{fig:lattice-thermal-conductivity}(a)].
This is in fact rather unsurprising since already elemental Ge has a rather mass variance due to multiple isotopes.

\subsubsection{Overview of experimental data}

The total thermal conductivity $ \kappa$ has been measured using a variety of techniques for both single and polycrystalline samples, see e.g., Refs.~\onlinecite{AviSueUme06a, TobChrIve08, SarSvePal06, ChrJohSon09}.
To extract the lattice contribution to the thermal conductivity, it is customary to remove the electronic part by the use of Wiedemann-Franz law.
Yet, as discussed in \sect{sect:electronic-thermal-conductivity}, the Wiedemann-Franz law is in fact only approximately valid under the relevant conditions.
For the sake of consistency, here we nonetheless use the data reported in the original papers and focus on single crystalline samples.

The compilation of the experimental results [\fig{fig:lattice-thermal-conductivity}(b)] illustrates a noticeable spread, especially at low temperatures.
The strong temperature dependence along with the pronounced low-temperature peak in the data for n-type \BGG{} \cite{ChrJohSon09, AviSueUme06a} indicate a crystal-like thermal conductivity down to very low temperatures.
By contrast, the much weaker temperature dependence and low-temperature plateau in the results for p-type material \cite{BenChrBry04, AviSueUme06a} are consistent with glass-like behavior.

This correlation between n-type (p-type) electrical and crystal-like (glass-like) thermal conductivity has been documented in Ref.~\onlinecite{TakSueNak14}.
The crossover between the two conductivity types occurs close to the stoichiometric composition of 16:30 with n-type (p-type) material being slightly Ga (Ge) deficient \cite{AviSueUme06a}.
It has been found experimentally \cite{ChrLocOve06} that Ba atoms, which are formally assigned to $6d$ Wyckoff sites, are less (more) displaced in n-type (p-type) material \cite{AngErh17}.

Interestingly, a numerical study of the low-frequency modes in \BGG{} and Ba$_8$Ga$_{16}$Sn$_{30}$ as cases for on and off-center Ba positioning has found crystal-like and glass-like conductivity, respectively \cite{LiuXiZho16}.
For the case of Ba$_8$Ga$_{16}$Sn$_{30}$ it has been furthermore argued that the low-temperature plateau is due to a delocalization-localization transition for the acoustic modes \cite{XiZhaChe17}.
The latter process in turn can only occur for off-center guest atoms, which induce disorder and level repulsion.
In this situation, the heat carrying quasi-particles become overdamped, i.e. the oscillation period is comparable to the lifetime, leading to a saturated and thus temperature independent lattice thermal conductivity \cite{XiZhaNak18}.
To properly capture this effect a non-perturbative treatment is required.
Here, we therefore limit ourselves to a comparison with experimental data for the thermal conductivity of n-type material.
We note, however, that the present approach for extracting IFCs can in principle be extended to the necessary higher expansion orders \cite{FraEriErh19}.

\subsubsection{Comparison between calculations and experiment}

The lattice thermal conductivities derived from IFCs obtained in the static limit by either PBE or vdW-DF-cx substantially underestimate the experimental data (\fig{fig:lattice-thermal-conductivity}, also see Ref.~\onlinecite{TadGohTsu15}).
This is a rather unusual observation as one more commonly finds calculations to overestimate the experimental data (see e.g., Ref.~\onlinecite{LinErh16}) as computational analysis commonly account only for some of the scattering mechanisms that are active in reality.

The behavior observed here can, however, be understood by considering the expression for the lattice thermal conductivity \eq{eq:thermal-conductivity} and the phonon dispersions (\fig{fig:ph_dispersions_dos}).
Since in the case of the static IFC calculations the rattler modes are located at lower frequencies than in the experimental data the avoided crossings with the acoustic modes \cite{ChrAbrChr08} occur at smaller $\vec{q}$ vectors and as a result the relative fraction of propagating modes that contributes to the thermal conductivity is reduced. 
This mechanism can also explain the lower thermal conductivity obtained from PBE relative to vdW-DF-cx calculations.

As shown above, one must account for the effect of phonon-phonon coupling on the frequency spectrum in order to obtain closer agreement with experiment (\fig{fig:ph_dispersions_dos}).
Accordingly, using temperature dependent IFCs (TDIFCs) in \eq{eq:thermal-conductivity} leads to a substantial increase in the thermal conductivity by a factor of two at 100\,K up to a factor of three at 600\,K.
This effect is crucial in order to achieve good agreement with the experimental data in this temperature range.
In addition, the temperature dependence of $\kappa_l$ shifts from $\kappa_l\propto T^{-1}$ to $\kappa_l\propto T^{-0.69}$, which rather closely follows the experimental trend.

The importance of temperature dependent frequencies is also evident from the calculations in Ref.~\onlinecite{TadTsu18}, in which the temperature dependence of the vibrational spectrum was included via the self-consistent phonon formalism based on a fourth-order model.
In the latter approach the frequencies are thus temperature dependent whereas the IFCs themselves are not.
In the present case, we employ (effective) TDIFCs up to third-order to achieve a very similar effect.\footnote{
  We do not include the data from Ref.~\onlinecite{TadTsu18} in \fig{fig:lattice-thermal-conductivity} since it is based on a different exchange-correlation functional (PBE) and a different model for the distribution of Ga and Ge over the host lattice sites, which would lead to a misleading comparison.
}
The present methodology moreover allows us to analyze the contribution from second and third-order TDIFCs to the thermal conductivity separately.
To this end, we also computed the thermal conductivity using second-order TDIFCs and static (zero-K) third-order IFCs.
The results (see Figure 1 of the Supplementary Information) demonstrate the (effective) temperature dependence to be important for both second and third-order IFCs.
Whereas including (only) second-order TDIFCs causes a change in the structure of the lifetimes as a function of frequency, including third-order TDIFCs further leads to a systematic increase in the life times by a factor of up to approximately two.

\subsubsection{Microscopic contributions}

\begin{figure}
    \includegraphics[scale=\myscale, center]{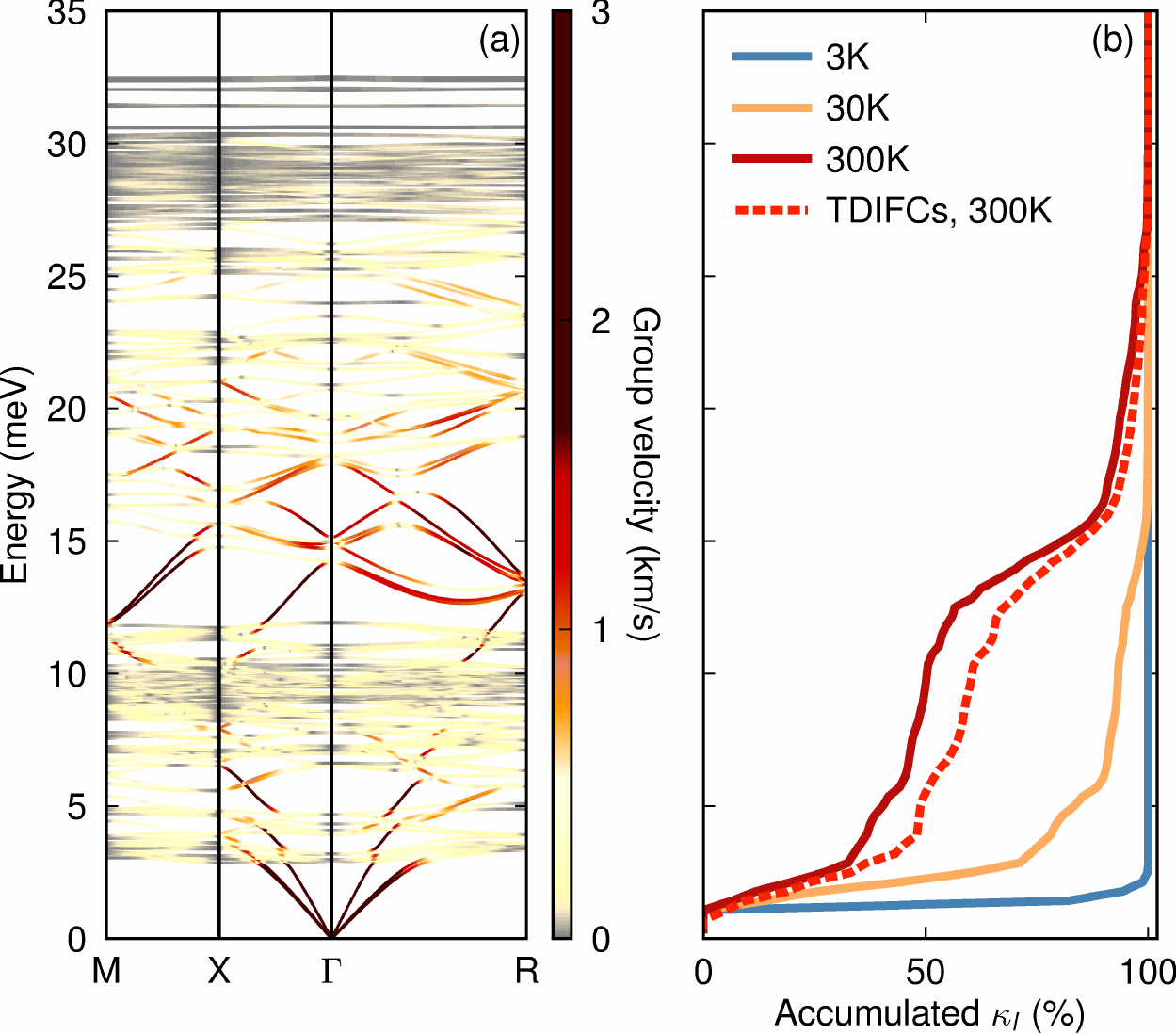}

    \vspace{0.3cm}

    \includegraphics[scale=\myscale, left]{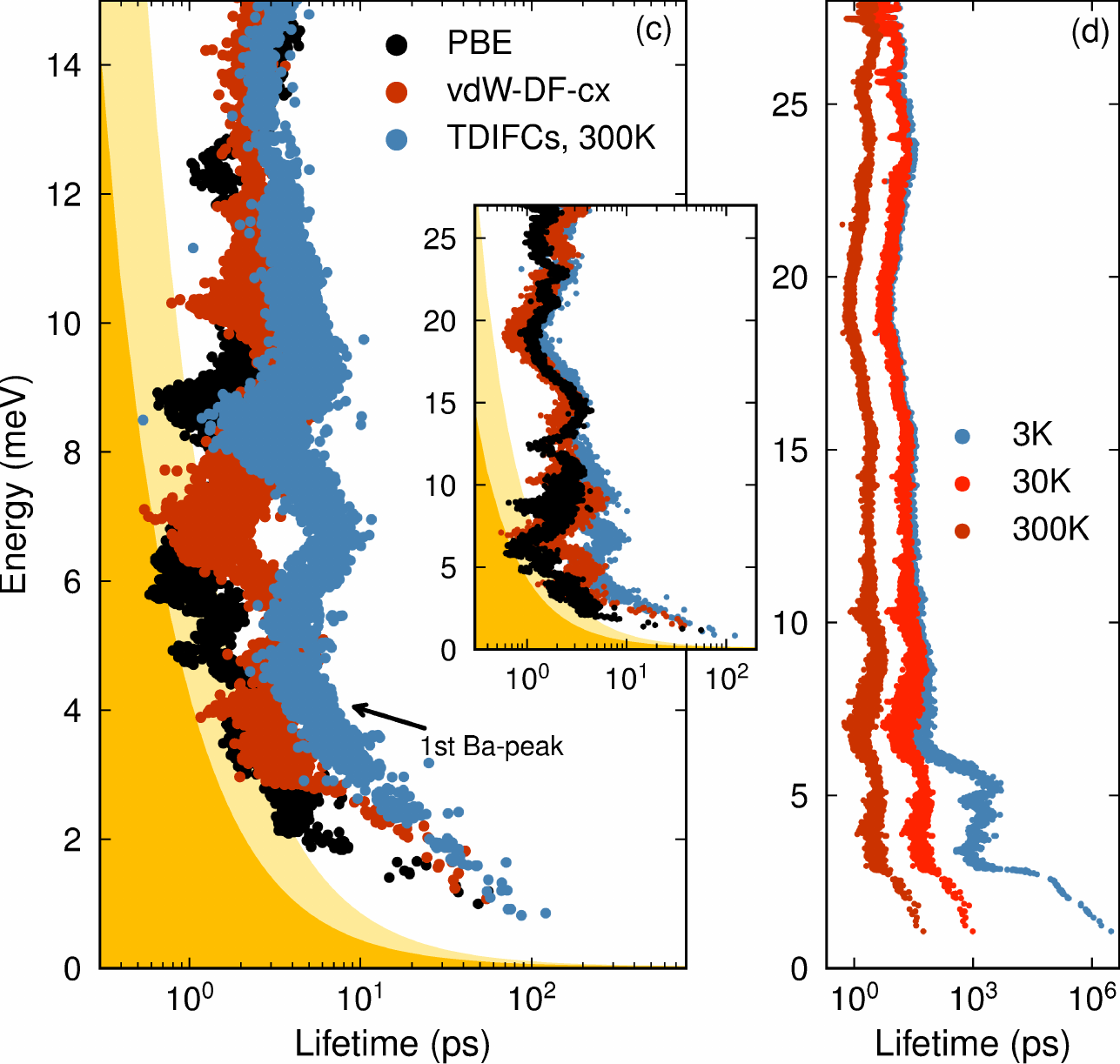}
    \caption{
        (a) Phonon dispersion in \BGG{} from static vdW-DF-cx IFCs.
        (b) Normalized accumulated lattice thermal conductivity with respect to energy as computed with static vdW-DF-cx IFCs at 3, 30 and \unit[300]{K} (solid lines), compared to calculations based on the temperature dependent force constants (TDIFCs) at 300\,K.
        (c) Phonon-phonon limited lifetimes at \unit[300]{K} computed with PBE IFCs (black markers), vdW-DF-cx IFCs (red markers) and TDIFCs (blue markers).
        The filled yellow curves indicate the overdamped region for a classical harmonic oscillator.
        (d) Comparison of lifetimes computed from static vdW-DF-cx IFCs at 3, 30 and \unit[300]{K}.
    }
    \label{fig:group-velocities-lifetimes}
\end{figure}

Inorganic clathrates have repeatedly been shown to exhibit very low thermal conductivities \cite{CohNolFes99, AviSueUme06a, TobChrIve08, SarSvePal06, ChrJohSon09} and have been discussed as realizations of the so-called phonon glass-electron crystal (PGEC) concept \cite{Row05, TakSueNak14}.
Accordingly, the mechanisms that give rise to the very low thermal conductivity in clathrates have been scrutinized experimentally as well as by theory and simulation, see, e.g., Refs.~\onlinecite{DonSanMyl01, MadSan05, ChrAbrChr08, MadKatBer16, TadGohTsu15, LiuXiZho16, YanCheHu17, LorPaiGio17, EucPaiGio18, TadTsu18}.
The primary object of attention has been the phonon dispersion \cite{MadSan05, ChrAbrChr08, LiuXiZho16}, which provides information about vibrational frequencies and group velocities and is more readily accessible both in experiments and calculations.
The phonon lifetimes in these materials have only been recently addressed using an approximate lifetime model for \BGS{} \cite{MadKatBer16} as well as perturbation theory in the case \BGG{} \cite{TadGohTsu15, TadTsu18, TadGohTsu15}.

The present calculations allow us to scrutinize the individual contributions to the lattice thermal conductivity according to \eq{eq:thermal-conductivity}, including group velocities and lifetimes, and thereby gain further insight into the glass-like thermal conduction in inorganic clathrates.
From the phonon dispersion [\fig{fig:group-velocities-lifetimes}(a)] two energy regions $ \omega <$ \unit[3]{meV} and \unit[12]{meV} $ < \hbar\omega <$ \unit[16]{meV} can be identified with large group velocities.
At low temperatures, the contribution to $\kappa_L$ stems mainly from the region $\omega <$ \unit[2]{meV} [\fig{fig:group-velocities-lifetimes}(b)].
Yet already at about 300\,K almost half of the heat transport is accomplished by modes with frequencies above \unit[3]{meV}.
This behavior is the result of two concurrent processes:
(\emph{i}) from about 300\,K all modes are occupied and the mode specific heat capacity \eq{eq:heat-capacity} saturates;
(\emph{ii}) the lifetimes of the acoustic modes in the lower energy window drop substantially whereas the lifetimes of the higher lying modes are relatively less affected [\fig{fig:group-velocities-lifetimes}(d)].
Compared to previous studies, which focused primarily on the low energy region, the present results thus demonstrate that at least in the case of \BGG{} important contributions stem from higher energy modes and must be included in order to obtain a sound description of the thermal conductivity.

When considering the static IFCs it appears that a large number of modes is actual either strongly damped or even overdamped [\fig{fig:group-velocities-lifetimes}(c)].
The oscillation period of these modes is thus comparable to their lifetime.
In the case of the TDIFCs the lifetimes are notably longer, which reflects the effective (albeit not formal) renormalization of the modes.

Overall, in agreement with previous work the present calculations demonstrate that weak coupling between host and low-lying guest (Ba) modes gives rise to avoided band crossings in the phonon dispersion \cite{MadSan05, ChrAbrChr08}, which in turn cause a dramatic reduction in the group velocities of almost all modes with frequencies above the lowest guest mode \cite{MadSan05, ChrAbrChr08}.
The few dispersed modes above this threshold are strongly damped and accordingly do not contribute notably to $\kappa_l$.
The very small thermal conductivity is thus the result of the extremely small Brillouin zone volume available to propagating phonon modes.
It should also be noted that since the effective mean free path drops to about 1\,nm already at 300\,K, further reduction of $\kappa_l$ by e.g., microstructural engineering \cite{BisHeBlu12} does not appear to be very promising.
It should also be recalled that in the temperature range relevant for thermoelectric applications, the electronic contribution $\kappa_e$ is already comparable to if not larger than $\kappa_l$ (see \sect{sect:electronic-thermal-conductivity}).

\subsection{Thermal conductivity: electronic contribution}
\label{sect:electronic-thermal-conductivity}

\begin{figure}
  \centering
  \includegraphics[scale=\myscale]{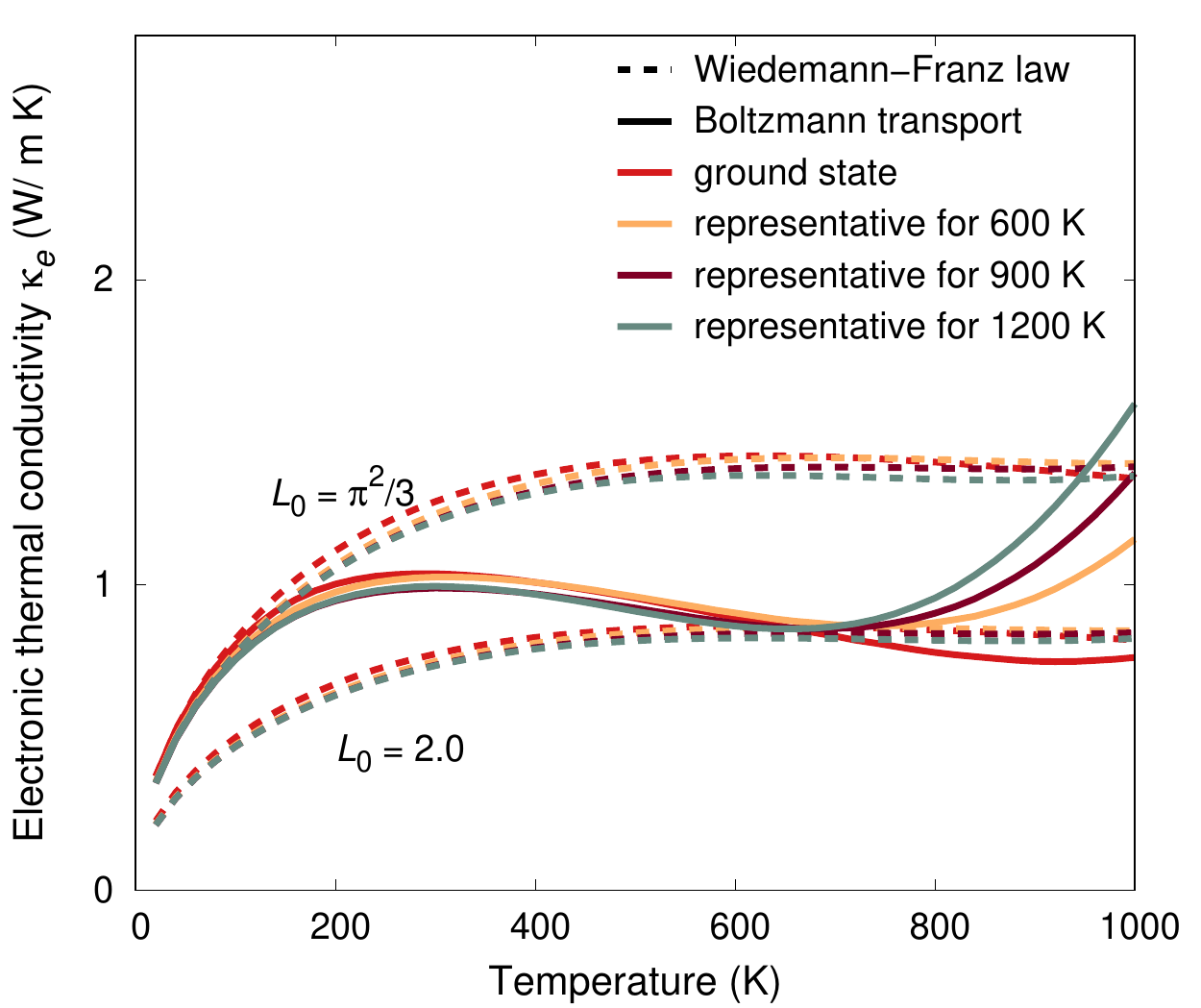}
  \caption{
    Electronic contribution $ \kappa_e$ to the thermal conductivity calculated for the ground state structure as well as for structures extracted from Monte Carlo simulations \cite{AngLinErh16} representative of the chemical order at different temperatures.
    Data obtained using the Wiedemann-Franz law $ \kappa_e=L\sigma T$ are shown by dashed lines, whereas the thermal conductivity obtained within the framework of Boltzmann transport theory \cite{SchAmbTho03, MadSin06} is shown by solid lines.
    Note that below 900\,K the two sets of data deviate by as much as 25\%, whereas above approximately 900\,K the BTT data indicates a sharp rise which is not predicted by the Wiedemann-Franz law.
  }\label{fig:electronic-thermal-conductivity}
\end{figure}

When conducting experiments only the total thermal conductivity $\kappa$ is directly accessible.
To resolve each contribution one therefore commonly resorts to the Wiedemann-Franz law.
The Wiedemann-Franz law couples the electrical conductivity $\sigma$ to the electronic thermal conductivity
\begin{align}\label{eq:Wiedemann-Franz-law}
    \kappa_e = L \sigma T,
\end{align}
by which the lattice thermal conductivity $\kappa_l$ is estimated.
Here, $L=L_0(k_B^2/e^2)$ is the Lorenz number.
When resolving experimental data it is common to use $L_0 = \pi^2/3 \approx 3.3$, which is obtained for a degenerate electron gas, or not specify the value of $L_0$ used.
As noted, e.g., in Ref.~\onlinecite{MahBar99} a value of $L_0=2$ should be used for a degenerate semiconductor (such as a typical thermoelectric clathrate).
Even then the Wiedemann-Franz law ought to be considered a low level approximation to the actual behavior as $L_0$ is not a universal constant.

We therefore conclude our analysis of the thermal conductivity in the prototypical inorganic clathrate \BGG{} by considering the electronic contribution $\kappa_e$ (\fig{fig:electronic-thermal-conductivity}).
At low temperatures $\lesssim\,200\,\text{K}$ the thus obtained $\kappa_e$ agrees rather well with the Wiedemann-Franz law assuming a degenerate electron gas ($L_0\approx 3.3$), whereas at higher temperatures $\gtrsim\,400\text{K}$ the semiconductor value $L_0=2$ yields better agreement.

Above approximately \unit[900]{K} the $\kappa_e$ from Boltzmann transport theory reveals a sharp rise for some structural models.
In this context one should note that the integrand in the expression for the electronic thermal conductivity \cite{SchAmbTho03, MadSin06} \eq{eq:electronic-thermal-conductivity} includes a term $ {(\epsilon_{i\vec{k}} - \mu_e)}^2$.
As a result, $ \kappa_e$ is most sensitive to contributions from states about $k_B T$ above and below the Fermi level $\mu_e$, rather than to states in the immediate vicinity of $\mu_e$, which dominate in the case of $ \sigma$.
The deviation at higher temperatures could, therefore, be an indication for shortcomings of the effective lifetime model, which does not distinguish these states.
In any case, the present analysis suggests that the Wiedemann-Franz law should be applied with caution when trying to discriminate the electronic and lattice thermal conductivities, and that a value of $L_0$ corresponding to a degenerate semiconductor is more appropriate for describing the situation in inorganic clathrates.

\section{Conclusions}
\label{sect:conclusions}

The very low thermal conductivities observed in inorganic clathrates are challenging to address both experimentally and computationally.
In the present study, focusing on \BGG, we have undertaken a systematic computational analysis of the various mechanisms and features that contribute to this property.

Firstly, we have addressed the role of the exchange-correlation functional in describing both structure and vibrational spectra, from which we concluded that the vdW-DF-cx method provides a well balanced description of inorganic clathrates.
Next by using temperature dependent interatomic force constants we demonstrated that phonon-phonon coupling (and thus temperature) must be taken into account in order to accurately capture the frequencies of the rattler modes, which also allowed us to predict correctly the experimentally observed temperature dependence of these modes.
On the other hand, the composition dependence of the rattler mode frequencies was shown to be small.

Based on this level of understanding we then predicted both the lattice and the electronic thermal conductivity.
For the former we obtained very good agreement with experiments using temperature dependent IFCs whereas we observed a pronounced underestimation when using IFCs representing the static limit.
The $\kappa_l$ values obtained using different IFCs could be rationalized by considering the relative Brillouin zone volume of propagating (heat carrying) modes.
Specifically, the underestimation of the rattler modes is associated with the onset of avoided crossings at a lower $\vec{q}$ vector.
The analysis furthermore reveals that IFCs obtained in the static limit yield heavily damped as well as overdamped quasi-particles, which is suggestive of glass-like transport.
Taking into account phonon-phonon coupling via temperature dependent IFCs (and thus effectively mimicking renormalization) leads to larger lifetimes and more well defined quasi-particles.

Finally, a comparison of predictions for the electronic contribution to the thermal conductivity demonstrates the Wiedemann-Franz law must be applied with more care when separating experimental thermal conductivity data as the $L_0$ pre-factor can vary between 2 (degenerate semi-conductor) and 3.3 (free electron gas) across the temperature range of interest.

The present results provide a very detailed perspective on the thermal conductivity in inorganic clathrates and shed light on the application of Boltzmann transport theory for predicting systems with strongly damped quasi-particles.
The thermal conductivity in clathrates has been studied in a few previous studies, including \BGG{} \cite{TadGohTsu15, TadTsu18}, pure silicon clathrates M$_{8-x}$Si$_{46}$ \cite{YanCheHu17, EucPaiGio18}, and the ordered clathrate Ba$_8$Au$_6$Ge$_{46}$ \cite{LorPaiGio17}.
With the exception of Ref.~\onlinecite{TadTsu18}, these studies employed IFCs obtained at zero K and restricted the range of interactions.
Here, we emphasized the importance of taking into account the effect of phonon-phonon coupling on the vibrational spectrum.
Furthermore, we demonstrated the efficacy of advanced regression schemes \cite{EriFraErh19} for extracting IFCs with minimal restrictions.

\section*{Acknowledgments}

This work was funded by the Knut and Alice Wallenberg Foundation, the Swedish Research Council as well as the Danish Council for Strategic Research via the Programme Commission on Sustainable Energy and Environment through sponsoring of the project ``CTEC -- Center for Thermoelectric Energy Conversion'' (project no. 1305-00002B).
Com\-puter time allocations by the Swedish National Infrastructure for Computing at NSC (Link\"oping) and PDC (Stockholm) are gratefully acknowledged.

\end{document}